\documentclass[12pt]{article}

\usepackage{graphicx}
\usepackage{epstopdf}
\usepackage{amsmath}
\usepackage{amsfonts}
\usepackage{amssymb}
\usepackage{enumerate}
\usepackage{amsthm}
\usepackage{array}
\usepackage{epsfig}
\usepackage[small,bf]{caption}
\usepackage{appendix}
\usepackage{dsfont}
\usepackage{hyperref}
\def\1ad{\mbox{\normalsize $^1$}}
\def\2ad{\mbox{\normalsize $^2$}}
\def\3ad{\mbox{\normalsize $^3$}}
\def\4ad{\mbox{\normalsize $^4$}}
\def\5ad{\mbox{\normalsize $^5$}}
\def\6ad{\mbox{\normalsize $^6$}}
\def\7ad{\mbox{\normalsize $^7$}}
\def\8ad{\mbox{\normalsize $^8$}}

\def\beq{\begin{equation}}                     %
\def\eeq{\end{equation}}                       %
\def\bea{\begin{eqnarray}}                     
\def\eea{\end{eqnarray}}                       
\def\dj{\hbox{d\kern-0.347em \vrule width 0.3em height 1.252ex depth
-1.21ex \kern 0.051em}}

\def\half{{1\over 2}\,}

\def\ket{\rangle}
\def\bra{\langle}

\def\pt{\partial}

\def\shalf{{\mbox{$\half$}}}

\def\Dirac{\,\raise.15ex\hbox{/}\mkern-13.5mu D}
\def\dirac{\,\raise.15ex\hbox{/}\kern-.57em \partial}
\def\pslash{\,\raise.15ex\hbox{/}\kern-.57em p}

\def\vep{\varepsilon}

\begin{document}

                     %

\newcommand{\sheptitle} {Holographic entanglement entropy probes (non)locality} \newcommand{\shepauthora} {{\sc Jos\'e
    L.F.~Barb\'on and Carlos A. Fuertes}}

\newcommand{\shepaddressa} {\sl
  Instituto de F\'{\i}sica Te\'orica  IFTE UAM/CSIC \\
  Facultad de Ciencias C-XVI \\
  C.U. Cantoblanco, E-28049 Madrid, Spain\\
  {\tt jose.barbon@uam.es}, {\tt carlos.fuertes@uam.es} }

\newcommand{\shepabstract} {
  \noindent

  We study the short-distance structure of geometric entanglement entropy in certain theories with a built-in scale of
  nonlocality. In particular we examine the cases of Little String Theory and Noncommutative Yang--Mills theory, using
  their AdS/CFT descriptions. We compute the entanglement entropy via the holographic ansatz of Ryu and Takayanagi to
  conclude that the area law is violated at distance scales that sample the nonlocality of these models, being replaced
  by an extensive {\it volume law}.  In the case of the noncommutative model, the critical length scale that reveals the
  area/volume law transition is strongly affected by UV/IR mixing effects. We  also present  an argument showing that
  Lorentz symmetry  tends to protect the area law for theories with field-theoretical density of states. }

\begin{titlepage}
  \begin{flushright}
    {IFTE UAM-CSIC/2008-17\\
    }

\end{flushright}
\vspace{0.5in} \vspace{0.5in}
\begin{center} {\large{\bf \sheptitle}}
  \bigskip\bigskip \\ \shepauthora \\ \mbox{} \\ {\it \shepaddressa} \\
  \vspace{0.2in}

  {\bf Abstract} \bigskip \end{center} \setcounter{page}{0} \shepabstract
\vspace{2.7in}
\begin{flushleft}
  \today
\end{flushleft}


\end{titlepage}

\newpage


\setcounter{equation}{0}

\section{\label{intro} Introduction}

\noindent

Entanglement entropy is a measure of quantum correlations in a bipartite decomposition of a quantum system.  If the
total Hilbert space admits a decomposition ${\cal H} = {\cal H}_A \otimes {\cal H}_B$ any state $\rho$ of the system
defines a reduced density matrix, $\rho_A$, for observables that are `blind' to, say ${\cal H}_B$, by simply tracing
over the degrees of freedom in ${\cal H}_B$.  Then, the entanglement entropy is defined as the von Neumann entropy of
this reduced density matrix:
\begin{equation}
  S_{A|B} = - \text{Tr} \,\rho_A \log \rho_A \quad , \qquad \rho_A = \text{Tr}_B \,\rho  \quad.
\end{equation}
For the case that $\rho = |\psi \ket \bra \psi |$ is a pure state, there is the same amount of entanglement in the two
parts of the system: $S_{A} = S_{B}$.

More specifically, one can adapt the bipartite decomposition of the Hilbert space to a certain basis of {\it localized}
degrees of freedom in regions of space $A$ and $B$, so that $A\cup B$ is the whole configuration space of the system.
This particular avatar of the entanglement entropy is also called {\it geometric entropy} and is a natural observable in
quantum theories with elementary degrees of freedom defined locally in space, such as lattice models and their idealized
long-distance descriptions as quantum field theories (QFT). In this paper, we discuss geometric entropy but keep using
loosely the terminology of `entanglement entropy'.

It can be argued on general grounds that geometric entropy in the {\it vacuum} state of a weakly coupled quantum field
theory satisfies the so-called \textit{area law}, i.e.~the entanglement entropy is proportional to the volume of the
boundary of the region under consideration, measured in units of an appropriate ultraviolet (UV) cutoff
\cite{enten}.\footnote{In fact it was this property that originally raised attention, because of its similitude to the
  entropy of black holes.}  In $d$ spatial dimensions one finds \beq\label{arealaw} S[A] \propto N_{\rm eff} \,{|\pt
  {A}| \over \varepsilon^{\;d-1}} + \dots \,,\eeq where $|\pt {A} | \equiv {\rm Vol} (\pt A)$ is the volume of the
$(d-1)$-dimensional boundary of $A$ and $N_{\rm eff}$ is the effective number of on-shell degrees of freedom (flavour,
spin, color).  The dots in (\ref{arealaw}) stand for subleading corrections in the short-distance expansion. Keeping the
finite terms in the continuum limit one can define renormalized versions of the entanglement entropy, whose structure
encodes properties related to physical energy thresholds like mass gaps \cite{nosotros}, confinement scales
\cite{kutkleb, nishitaka}, etc.  Corrections  to (\ref{arealaw}) are also important when considering entanglement in
non-vacuum sectors (see for example \cite{sdas}), such as thermal states.
 
More recently entanglement entropy has emerged as a useful order parameter of different phases with nonlocal quantum
order, particularly in the context of quantum phase transitions at zero temperature (cf. \cite{latorre}) and systems
with so-called topological order (cf. \cite{preskill}).

The area law (\ref{arealaw}) is strictly violated in the case of $(1+1)$-dimensional critical points.  In this case one
finds a logarithmic behavior 
\beq\label{dosdcrit} S[A]_{d=1} = {c \over 3} \,\log\, ( | A | / \varepsilon )\;, \eeq where
$|A | = {\rm Vol} (A)$ is the volume (length in this case) of the region $A$. The central charge $c\sim N_{\rm eff} $
arises with a universal coefficient, even in the case of strongly coupled CFTs.  The area law does apply away from the
critical point, i.e.~for small correlation length $\xi < | A |$, although the logarithmic cutoff dependence still
persists,\footnote{In $d=1$, $|\partial A|$ stands for  the number of boundary points.}
\begin{equation}
  S[A]_{d=1} = \frac{c}{6} \,|\pt A |\, \log \frac{\xi}{\varepsilon}  \;.
\end{equation}

Beyond this two-dimensional `anomalous' behavior, one can associate the area law (\ref{arealaw}) with {\it local} QFTs
defined in terms of UV fixed points. This is even the case at strong coupling, at least for those UV fixed points that
can be studied via the AdS/CFT correspondence \cite{adscft}.   See \cite{cirac, eisert} for a recent discussion of the generality of the area law in lattice systems.  On the other hand, a {\it volume law} of
the entanglement entropy can be associated to a violation of locality in the underlying theory.  In order to argue this
point at a heuristic level, we can consider a nonlocal version of the Heisenberg's antiferromagnetic spin chain,
\begin{equation}
  H = J \sum_{\langle i,j \rangle} \mathbf{S}_i\cdot \mathbf{S}_j\;,
\end{equation}
where $J>0$ and the sum runs over pairs of spins chosen uniformly at random, in such a way that each spin belongs to
only one pair.  The ground state is then the direct product of singlets
\begin{equation}
  \frac{1}{\sqrt{2}} \left( | \uparrow_i \downarrow_j \rangle - | \downarrow_i \uparrow_j \rangle \right) \quad,
\end{equation}
for each pair of sites. Each singlet contributes $\log 2$ to the entanglement entropy when the spins sit on opposite
sides of the boundary, and zero in all other cases.  Hence, the entropy is proportional to the number of singlets
connecting the `inside' and the `outside'.  For the model at hand, this is on average just the number of spins found
inside $A$, i.e.~$S[A] \sim |A| / \varepsilon$, a volume law.

In this paper we provide further evidence linking the {\it extensivity} of the entanglement entropy with nonlocal
behavior in the underlying theory.  More specifically we study the examples of Little String Theory (LST) and
noncommutative Yang--Mills Theory (NCYM) (see \cite{aharonyrev, douglasnek} for reviews with a collection of early
references on these subjects), using the AdS/CFT ansatz \cite{japoneses} for the entanglement entropy in the holographic
description of these models.  We find that the volume law \beq\label{vollaw} S[A] \propto {|A| \over \varepsilon^{\,d}}
\eeq takes over the area law (\ref{arealaw}) when the characteristic size of $A$, defined as \beq\label{sizel} \ell
\equiv 2{|A| \over |\pt A |} \, \eeq falls well below the critical nonlocality length, $\ell \ll \ell_c$. It is
important to emphasize that we are referring here to extensivity of the leading short-distance term in the entanglement
entropy, rather than the finite, cutoff-independent terms that can be identified as subleading corrections to
(\ref{arealaw}). These UV-finite terms are quite interesting and the subject of some recent attention
(cf.~\cite{japoneses, kutkleb, nosotros}) but will not be the main subject of this paper.

This paper is organized as follows. In section \ref{ads} we review the robustness of (\ref{arealaw}) for UV fixed points
in the AdS/CFT representation and provide some insight on this fact by examining non-conformal examples of strongly
coupled theories that can nevertheless be considered as local. In section \ref{sec:nonlocal} we study the entanglement
of nonlocal theories. In subsection \ref{sec:LST} we focus in the holographic description of LST and verify the
emergence of a volume law at short distances. In subsection \ref{sec:noncommutative} we do the same for the holographic
description of NCYM. In section \ref{sec:epilogue} we discuss how Lorentz symmetry at the boundary together with
standard density of states are sufficient to guarantee the area law. We end with some conclusions in section
\ref{sec:conclusions}.

\section{\label{ads}Holographic entanglement entropy and locality}

\noindent

The holographic ansatz for the calculation of entanglement entropy in theories with UV fixed points \cite{japoneses}
incorporates in a natural way the area law (\ref{arealaw}) (see also \cite{lista} for further developments).  Any such
holographic model is defined by a background of string or M-theory with asymptotic geometry ${\rm AdS}_{d+2} \times
K_{d_K}$ near the boundary, where $K_{d_K}$ is a compact Einstein manifold of dimension $d_K$. Away from the boundary
the geometry can be more complex and background fields of various types may be excited, representing the breakdown of
strict conformal symmetry by energy thresholds. We take the conformal boundary of the AdS$_{d+2}$ at infinity to be
given by a flat $(d+1)$-dimensional Minkowski space $\mathds{R}^{d+1}$. Let $A$ denote a purely spatial, $d$-dimensional
domain in $\mathds{R}^d$ with a smooth boundary $\pt A$ and let ${\overline A}$ denote the minimal $d$-dimensional
hypersurface in the bulk whose boundary on $\mathds{R}^{d+1}$ precisely coincides with $\pt A$. Then, the holographic
ansatz for the entanglement entropy is \beq\label{holoansatz} S[A] = {{\rm Vol} (\,{\overline A}\,) \over 4G}\;, \eeq
where $G$ is Newton's constant and the induced volume form of the bulk is defined in the Einstein frame. The
hypersurface ${\overline A}$ is of codimension two on the complete bulk spacetime of dimension $d+2+d_K$, and
furthermore completely wraps any compact internal cycle, such as the Einstein manifold $K_{d_K}$ that is visible
asymptotically. Hence, we can specify further (\ref{holoansatz}) by working with the Kaluza--Klein reduction to $d+2$
dimensions and taking $G=G_{d+2}$ as the induced Newton's constant. Alternatively, in the particular examples of this
paper we will mostly deal with ten-dimensional backgrounds of type II string theory, and we may as well work in
string-frame variables with an explicit dilaton background:
\begin{equation} \label{eq:ent_ent_2} S[A ]= \frac{1}{32\pi^6 \alpha'^4}\int_{\bar{A}} d^8\sigma\,
  e^{-2\phi}\sqrt{G_{\mathrm{ind}}^{(8)}} \quad,
\end{equation}
where $G_{\mathrm{ind}}^{(8)}$ denotes the determinant of the string-frame induced metric into $\bar{A}$ from the bulk,
and the dilaton $\phi$ is normalized so that the local value of the ten-dimensional Newton's constant is $G_{10} (\phi)
= 8\pi^6 \alpha'^4 e^{2\phi}$.

We can now give a simple heuristic argument that explains the universality of (\ref{arealaw}) in any holographic
background asymptotic to an AdS$_{d+2}$ spacetime with metric \beq\label{adshol} ds^2 \longrightarrow R^2 \,u^2
\left(-dt^2 + dx_d^2 \,\right) + R^2 {du^2 \over u^2}\;, \eeq where $R$ is the AdS radius of curvature. With this choice
of coordinates, the holographic variable $u$ has dimensions of energy and directly represents a fiducial energy scale
parameter in the dual CFT. The conformal symmetry of the dual CFT is characterized by the scaling invariance of the bulk
metric (\ref{adshol}) under the combined transformation $(t,x_d, u) \rightarrow (\lambda\, t, \lambda \,x_d, \lambda^{-1}
u)$.  A minimal $d$-surface ${\overline A}$ in AdS$_{d+2}$ with boundary $\partial A$ penetrates into the bulk down to a
`turning point' $u\sim u_*$. Conformal symmetry implies that the minimization problem has no intrinsic length scale (the
overall AdS radius $R$ drops out of the variational problem).  Therefore, using (\ref{sizel}) as a measure of the size of
$A$, we must have $u_* \sim 1/\ell$, provided $u_*$ still remains well within the region where the AdS metric
(\ref{adshol}) is a good approximation. The minimal surface is locally a cylinder of the form $\pt A \times
[u_\varepsilon , \infty]$ near the boundary, so that its volume gets a cutoff-dependent contribution of the form \beq
\label{uvdiv} {\rm Vol} (A)_{\rm UV} \sim R^{d} \,|\,\pt A\,| \int^{u_\varepsilon} {du\over u} \, u^{d-1} \sim R^{d}
\,|\,\pt A\,| \,{u_\varepsilon^{d-1} \over d-1}\;, \eeq which reproduces (\ref{arealaw}) with $u_\varepsilon \sim
\varepsilon^{-1}$, since $N_{\rm eff} \sim R^{d} / G_{d+2} $ according to the standard AdS/CFT dictionary.

Heuristically we can associate the locality of the theory to the occurrence of a UV/IR relation of `Heisenberg' type:
$\ell(u_*) \sim 1/u_*$, since the radial coordinate $u$ is interpreted as an energy scale of the CFT.  We will regard
this relation as the `footprint' of a local theory, even in cases where the conformal symmetry is strongly violated.

An interesting example is provided by all the theories arising as holographic duals of D$p$-brane backgrounds in type II
string theory, i.e.~super Yang--Mills models in $p+1$ dimensions, with gauge group $SU(N)$, and (dimensionful) 't Hooft
coupling parameter $\lambda = g^2_{\rm YM} N$ (cf. \cite{itzhaki}). The relevant string-frame metric is scaled at the
near-horizon region of the D$p$-brane backgrounds: \beq\label{dpes} ds^2 /\lambda^{1\over 5-p} \propto u^{7-p \over 5-p}
\left(-dt^2 + dx_p^{\,2} \right) + u^{p-3 \over 5-p} \left({du^2 \over u^2} + d\Omega_{8-p}^2 \right)\;, \eeq in units
$\alpha' =1$, and the dilaton profile \beq\label{dilp} e^{-2\phi} \propto N^2 \,\lambda^{p-7 \over 5-p} \,u^{(7-p)(3-p)
  \over 5-p} \;, \eeq generalizing the conformal $p=3$ case. We use the radial energy variable $u$ introduced in \cite{peetpol} and we neglect $O(1)$ numerical constants for the purposes of
this discussion.  Furthermore, it will be enough to estimate the entropy over trial cylinders capped at $u=u_*$,
resulting in an expression \beq\label{trial} S[A] \sim N^2 \,|A|\,\lambda^{p-3 \over 5-p} \,u_*^{\,{9-p \over 5-p}} + {5-p \over 4}\,
N^2\,\lambda^{p-3 \over 5-p} \,|\pt A | \,\left( u_\varepsilon^{\,{4 \over 5-p}} - u_*^{\,{4 \over 5-p}}\right) \;, \eeq
which is extremal at the same Heisenberg-like UV/IR relation that featured in the conformal case: \beq\label{heisp} u_*
\sim {|\pt A| \over 2|A|} \equiv {1\over \ell}\;.  \eeq For $p<5$ this extremal surface is actually a local minimum of
the entropy functional (\ref{dpes}) and the resulting entanglement entropy scales as \beq\label{dpeen} S[\ell] \sim
N_{\rm eff} (\varepsilon) {|\partial A| \over \varepsilon^{\,p-1}} - C_p\, N_{\rm eff} (\ell) {|\partial A| \over
  \ell^{\,p-1}}\;, \eeq with $C_p $ an $O(1)$ numerical constant. We find a local `area law' with a renormalized
effective number of degrees of freedom \footnote{A similar result can be obtained for models with a logarithmic
  deviation from a fixed point, such as the gravity duals of `cascading gauge theories' \cite{klebstr}, where $N_{\rm
    eff}$ shows a logarithmic growth at high energies (cf. \cite{kutkleb}).}  \beq\label{neff} N_{\rm eff} (\varepsilon)
= N^2 \left({\lambda \over \varepsilon^{p-3} } \right)^{p-3 \over 5-p} \;.  \eeq This growing number of degrees of
freedom with energy is the same that becomes exposed when we excite the high-energy sector of the theory by thermal
states. Here, a natural definition is to measure the effective number of degrees of freedom in terms of the thermal
entropy density in units of the temperature of the system. In the bulk description, we estimate the thermal entropy
density $s(T)$ by that of black holes in the background (\ref{dpes}), according to the generalized AdS/CFT rules. The
result is (cf. \cite{itzhaki}) \beq\label{termeff} N_{\rm eff} (T) \equiv {s(T) \over T^p} \sim N^2 \left( \lambda
  T^{\;p-3} \right)^{p-3 \over 5-p} \;.\eeq Hence, the degrees of freedom that are being measured by the entanglement
entropy in the UV are the same degrees of freedom that account for the entropy of a Yang--Mills plasma at strong
coupling.
  
The discussion of D$p$-brane systems must be restricted to the regime where the effective dimensionless 't Hooft
coupling $\lambda_{\rm eff} \sim \lambda T^{\,p-3}$ is very large, since this is the regime where the geometry is
appropriately weakly curved. At the same time, $N$ must be large enough so that the string loop expansion is under
control. Beyond these thresholds one must use a variety of dualities to map out the phases of the system (cf.  for
example \cite{itzhaki, kogant}).

More fundamental is the restriction to $p<5$. At $p=5$ the previous formulas clearly break down, with $N_{\rm eff}$
becoming formally infinite, suggesting that the dual theory has a tower of field-theoretical excitations (a string
theory). We will address this case in the next section, as our first example of a nonlocal theory.  For $p>5$ there are
no working examples of holography (for example, the density of states of black holes leads to negative specific
heat). At the level of the previous formulas, the minimal hypersurface is pushed all the way to the cutoff scale $u_* = u_\vep$, a first example of a {\it volume law}, albeit somewhat pathological (see section 4 for a thorough discussion of these cases). 

\section{Nonlocal Theories}
\label{sec:nonlocal}
\noindent

In what follows, we turn to two examples of theories with an IR fixed point, i.e.~a CFT limit at low energies, but with
a built-in scale of nonlocality. In the dual geometrical description, we have backgrounds which approach AdS at {\it
  low} values of the energy variable, $u$, but differ very significantly at the UV boundary.

We start with the gravity dual of the worldvolume theory of NS5-branes \cite{lindil}. This is related by an S-duality to
the marginal case of D5-branes referred to in the previous section. Since the holographic formula for the entanglement
entropy can be written in terms of the Einstein-frame metric, which is invariant under S-duality, the
conclusions can be transported between Neveu--Schwarz and Dirichlet type five-branes.

Therefore, our first example arises naturally as the borderline case from the point of view of the arguments in the
previous section. In particular, it corresponds to a formally infinite number of field-theoretical degrees of freedom
$N_{\rm eff} = \infty$. Not surprisingly, the dual system turns out to be a string theory, albeit of a very exotic
variety.

The second example is of a different nature. We examine noncommutative Yang--Mills theories (NCYM) using their
holographic description \cite{noncads}. In this case, it is known that the nonlocality is of a milder nature, since it
does not involve an infinite tower of field-theoretical degrees of freedom. Rather, it has to do with the violation of
the microcausality rules enforced by Lorentz invariance. Accordingly, $N_{\rm eff}$ plays a less decisive role in this
case, but nevertheless we will confirm that the entanglement entropy still probes the noncommutative nonlocality
exposing a {\it volume law} at short distances.

\subsection{Little String Theory}
\label{sec:LST}
\noindent

Little String Theory (LST) is defined as the decoupled world-volume theory on a stack of $N$ NS5-branes, in the limit
$g_s \rightarrow 0$ with fixed string slope $\alpha'$.  The effective length scale of the theory is the combination $R=
\sqrt{N\alpha'}$. For large values of the rank, $N$, we have a dual geometrical description in terms of the near-horizon
region of the NS5-branes background \cite{callanstro}:
\begin{equation} \label{eq:tube} ds^2 = -dt^2 + dx_5^2 + \frac{R^2}{r^2} dr^2 + R^2 d\Omega_3^2 \quad, \qquad e^{\phi} =
  \frac{g_s R}{r} \quad,
\end{equation}
where $(t, x_5) \in \mathds{R}^{1+5}$ parametrizes the NS5-branes world-volume, i.e. the spacetime of the LST. Changing
variables to $r= g_s R \exp(z/R)$ yields \beq\label{cilingro} ds^2 = -dt^2 + dx_5^2 + dz^2 + R^2 \,d\Omega_3^2\;, \qquad
\phi(z)= -{z\over R}\;, \eeq confirming that $R$ is the unique length scale of the problem, with a geometry
$\mathds{R}^{5+1} \times \mathds{R} \times {\rm S}^3$, the product of a fixed-radius sphere and a {\it flat} cylinder,
and a linear dilaton of slope $1/R$ (cf. \cite{lindil}). One can have type IIA and IIB NS5-branes giving rise to two
different LSTs. We will focus on the type IIA case which has a clear holographic dual.

\begin{figure}
  \label{tubedualities}
  \begin{center}
    \epsfig{file=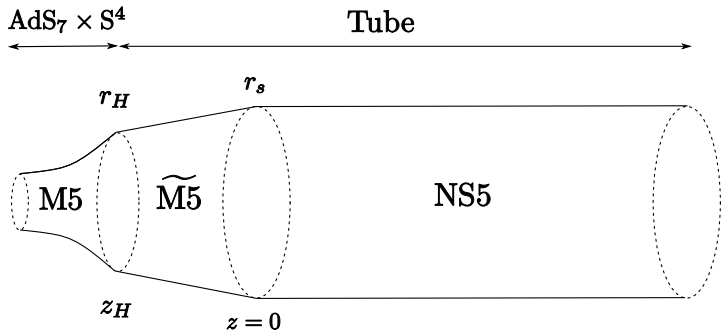, width= 9.5cm}
    \caption{ The different regions of the bulk type IIA background. The local string coupling grows towards smaller radii in the NS5 `tube', becoming of $O(1)$ at $r\sim r_s = g_s R$. At lower radii the system is well approximated by the uplifted solution to eleven dimensions, i.e.  the smeared ${\widetilde {\rm M5}}$-brane solution, which localizes below $r\sim r_H = g_s R /\sqrt{N}$ and flows to the ${\rm AdS}_7 \times {\bf S}^4$ dual of the $(2,0)$ CFT in six dimensions. In the type IIB case, the ${\widetilde {\rm M5}}$ phase
is replaced by the near-horizon D5-brane background, and the matching at $r\sim r_H$ takes the system to a non-geometrical phase described by weakly-coupled Yang--Mills theory.}
  \end{center}
\end{figure}

The interpretation of the dual theory as a string theory is borne out by the consideration of the density of
states. According to the most basic of holography rules, we expect the high-energy spectrum to be well approximated by
black holes in the background (\ref{cilingro}), which we will call `the tube' in what follows. Black solutions with
translational invariance on $\mathds{R}^5$ can be written down by the substitution $dt^2 \rightarrow h(z,z_0) \,dt^2$
and $dz^2 \rightarrow dz^2 /h(z,z_0)$, with the same dilaton profile and \beq\label{hfu} h(z,z_0) = 1-\exp\left(2(z_0
  -z)/R\right) \;.  \eeq These black holes have a {\it constant} intrinsic temperature $T_H = (2\pi R)^{-1}$,
independent of $z_0$, and moreover their Bekenstein--Hawking entropy yields a density of states of `Hagedorn' type,
betraying a stringy interpretation \cite{maldafive}: \beq\label{hages} \Omega(E)_{\rm BH} = \exp(E / T_H)\;, \qquad
E(z_0) = N E_H \exp(2z_0 /R)\;.  \eeq In this expression $E(z_0)$ is the energy of the LST state that corresponds to a
black hole with horizon at $z_0$. $E_H$ is a threshold energy defined as \beq\label{thsen} E_H = 2\pi \,N^3 \,V_5
\,T_H^6\;, \eeq corresponding to the internal energy of a six-dimensional gas of $N_{\rm eff} =N^3$ massless degrees of
freedom at temperature $T_H$. Hence, $E_H$ is the energy at which the LST matches to its low-energy limit, the $(2,0)$
six-dimensional CFT. In the holographic description, this matching occurs at $z=z_H = -R \log \sqrt{N}$, or $r= r_H = g_s R /\sqrt{N}$, and might be
regarded as the `infrared end' of the tube. For energies below $E_H$ the density of states is well approximated by that
of a six-dimensional CFT, with a holographic dual ${\rm AdS}_7 \times {\rm S}^4$ background of eleven-dimensional
supergravity. To be more precise (see for example \cite{itzhaki}), one finds the near-horizon limit of a stack of
M5-branes localized in a circle, with metric
\beq\label{loca}
ds^2 = H^{-1/3} (-dt^2 + dx_5^2 ) + H^{2/3} \left(dx_{11}^2 + dr^2 + r^2 d\Omega_3^2 \right)
\;,
\eeq
and profile function
\beq\label{defhache}
H(r) = \sum_{n\in \mathds{Z}} {\pi N \ell_p^3 \over \left[r^2 + (x_{11} - 2\pi R_{11} n )^2 \right]^{3/2}}\;,
\eeq
where the $11^{th}$ Planck length and circle radius are given by $\ell_p^3 = g_s \ell_s^3$, $R_{11} = g_s \ell_s$, with $\ell_s = \sqrt{\alpha'}$.
 Setting $\rho^2 = r^2 +
x_{11}^2$, this geometry is well approximated at $\rho \ll R_{11}$ by  
\begin{equation} \label{eq:nearHorizonM5}
    ds^2 \approx  \frac{\rho}{{(\pi N \ell_p^3)}^{\frac{1}{3}}} \left(-dt^2 + dx_5^2\right) + \frac{{(\pi N \ell_p^3)}^{\frac{2}{3}}}{\rho^2}  \left(d\rho^2 + \rho^2\, d\Omega_4^2\right)
\end{equation}
which adopts  the canonical ${\rm AdS}_7 \times {\rm S}^4$ form
under the change of variables $\rho = 4 \pi N \ell_p^3 u^2$, with $R_{\rm AdS} = 2R_{{\rm S}^4} = 2\ell_p (\pi N)^{1/3}$ and  $u$ the fiducial energy coordinate of the dual six-dimensional CFT. 

On the other hand, for $r_H \ll r \ll r_s \sim g_s R$ the sum in (\ref{defhache}) may be approximated by the first term alone, and we get the metric of $N$ M5-branes smeared over the $11^{th}$ circle. 
  In turn, this is nothing but the
$11^{th}$ dimensional  `uplift' of the tube geometry: \beq\label{uplift} ds_{11}^2 = e^{4\phi/3} dx_{11}^{\,2} +
e^{-2\phi/3} ds_{10}^{\,2} \;, \eeq with $ds_{10}^2$ and $\phi$ given by (\ref{cilingro}). At $z=0$ (or $r=r_s$)  the
$11^{th}$  circle acquires Planckian size, corresponding to the local string coupling of the type IIA
description becoming of order one.  The thermodynamic functions of black holes in these spaces
are independent of the uplifting operation, when expressed in terms of physical energy, entropy and temperature parameters. In practice, we can compute using (\ref{cilingro}) and extend analytically the results down to $z= z_H$, where one matches to the computations done with the metric
(\ref{eq:nearHorizonM5}). For this reason, we shall refer to the whole $z\geq z_H$ region as `the tube' in what follows.

\subsubsection{Entanglement entropy in the LST regime}

\noindent

Let us compute the entanglement entropy of a region of size $\ell$ in $\mathds{R}^5$, using (\ref{cilingro}) as bulk
geometry.  The precise formulas obtained can be readily extended to the eleven-dimensional intermediate regime in the
region $z_H < z < 0$, using the metric (\ref{uplift}), just as was the case for the thermodynamic functions.  This results from the fact that the eleven-dimensional bulk
hypersurface wraps the $x_{11}$ direction and the volume form of (\ref{uplift}) satisfies \beq\label{matcele} d{\rm
  Vol}_{11} = e^{-2\phi} dx_{11} \wedge d{\rm Vol}_{10} \;, \eeq so that both eleven-dimensional and ten-dimensional
formulae give the same basic integral for the entropy as a function of the boundary data at large $z$.  

For calculational convenience we will consider the particular case of the strip: $A = [-\ell /2, \ell /2] \times
\mathds{R}^{4}$. By translational symmetry on the $\mathds{R}^4$ factor, we can work in terms on the entropy density
$s[\ell]$ with the volume of $\mathds{R}^4$ factored out. We have a functional
\begin{equation}
  s[\ell]=  \frac{ | {\rm S}^3 |}{32\pi^6 \alpha'^4 g_s^2 R^2 }  \int_{-\frac{\ell}{2}}^{\frac{\ell}{2}} dx\, r^2
  \sqrt{ 1 + \frac{R^2}{r^2} \left({dr\over dx}\right)^2} \quad,
\end{equation}
where $|{\rm S}^3 |= R^3 \Omega_3$ is the volume of the 3-sphere. The bulk hypersurfaces are of the `straight belt'
form, $ {\overline A} = \mathds{R}^4 \times \gamma[r_*]$, where $\gamma[r_*]$ is a curve $r(x)$ subtending an asymptotic
length $\ell$ on the boundary as $x\rightarrow \pm \ell/2$ and turning at $r_* = r(0)$, defined by $\pt_x r(0) =0$.  The
smooth extremizing hypersurface verifies then
\begin{equation} \label{degl} \ell(r_*) = 2 R r_*^2 \int_{r_*}^{\infty} \frac{dr}{r \sqrt{r^4 - r_*^4}} = \frac{\pi}{2}
  R\;,
\end{equation}
a very peculiar result that was already obtained in Ref.~\cite{japoneses}. It shows that no smooth extremal surface
exists if the opening at the boundary is different from $\ell =\ell_c \equiv \pi R/2$.  Conversely, for $\ell = \ell_c$
there are an infinite number of them, parametrized by the turning point $r_*$. The entropy density at fixed $r_*$ is
\begin{equation}\label{entlst}
  s[r_*] = { \Omega_3 R^2\over 16\pi^6 g_s^2 \alpha'^4 }  \int_{r_*}^{r_\varepsilon} {r^3 dr \over \sqrt{r^4 - r_*^4}}
  =  {\Omega_3 R^2 \over 32\pi^6 g_s^2 \alpha'^4 } \sqrt{r_\varepsilon^4 - r_*^4}
  \;,
\end{equation}
where we have introduced $r_\varepsilon$ as a regularization cutoff.  This quantity is minimized for $r_* =
r_\varepsilon$, suggesting that the minimal surface degenerates at the UV cutoff.

In order to further interpret this situation we shall consider the approximate minimization problem for a restricted set
of hypersurfaces with the form of a cylinder of base $\pt A$ and extending down to $z=z_*$, in the coordinates of
(\ref{cilingro}). At $z=z_*$ we cap the cylinder with a copy of $A$. The contribution of the cylindrical part to the
entropy is \beq\label{cyll} S_{\rm cyl} = {1\over 32\pi^6 \alpha'^4} \int_{z_*}^{z_\varepsilon} dz \,e^{2z/R} \, |\pt A
| \, |{\rm S}^3 | = C\, N^4 \,T_H^{\,5} \,|\pt A |\,R \,\left(e^{2z_\varepsilon /R} - e^{2z_* /R}\right) \;, \eeq where
we have used the Hagedorn temperature $T_H = (2\pi R)^{-1}$ and defined the constant $C = \Omega_3 /2\pi$. The
contribution of the endcap is \beq\label{lstcap} S_{\rm cap} = {1\over 32\pi^6 \alpha'^4}\, e^{2z_* / R} \,| A | \,
|{\rm S}^3 | \;.  \eeq Combining the two, we have \beq\label{cilent} S[A] \sim C \,N^4\, T_H^{\,5} \,{R} \,|\pt A |
\,e^{2z_\varepsilon /R} + C\,N^4 \,T_H^{\,5} \, e^{2z_* /R} \left(2 |A| - R \, |\pt A| \right)\;.  \eeq With the
standard definition of the size of $A$, $\ell = 2|A| / |\pt A|$ we see that the minimal hypersurface within this
restricted class degenerates to $z_* = -\infty$ for $\ell > R$, or to $z_* = z_\varepsilon$ for $\ell < R$. In the
marginal case $\ell= R$ there is a degeneracy with respect to $z_*$, corresponding to the continuous degeneracy found in
(\ref{degl}), with a slightly renormalized value of the critical length, due to the non-smoothness of the class of
hypersurfaces considered here.

Hence, we find that the entropy satisfies a volume law at short distances.  We can interpret the cutoff factor
$\exp(2z_\varepsilon /R)$ in terms of LST physical quantities using Eq.~(\ref{hages}). Namely, if $E_\varepsilon$
denotes the energy of the largest black hole that fits inside the cut-off tube, then we have $ \exp (2z_\varepsilon /R)
= E_\varepsilon /N E_H$, and we can finally write down the volume law in the form \beq\label{volaf} S[A] \propto N_{\rm
  eff}(E_\vep) \, {|A| \over \ell_c^5} \;, \qquad {\rm for} \;\;\; |A| < \shalf \ell_c \,|\pt A|\;.  \eeq with an
effective cutoff length $\ell_c \sim 1/T_H \sim R$, and a running effective number of degrees of freedom given by
\beq\label{effnlst} N_{\rm eff} (E_\vep) = N^3 \,{E_\varepsilon \over E_H} \;, \eeq Just as in the case of D$p$-branes,
this effective number of degrees of freedom corresponds exactly to the effective number of thermally excited states
counted by a black hole of energy $E_\vep$. A very interesting aspect of (\ref{volaf}) is the treatment of the
ultraviolet cutoff. The landmark of locality, i.e. Heisenberg-like UV/IR relation, breaks down and yet we must implement
a cutoff procedure. The only way to enforce such a cutoff is in terms of the {\it total} energy of the system (see
\cite{littlehag} for a thorough discussion of this phenomenon in the context of LST thermodynamics).

\subsubsection{Infrared matching}
\noindent

The behavior for $\ell >\ell_c$ cannot be read off directly from (\ref{eq:tube}), since we know that the `tube' ends at
$z_H = -R \log \sqrt{N}$ and we have to match the geometry to the near-horizon limit of a stack of M5-branes, the dual
of a six-dimensional conformal field theory with $N_{\rm eff} = N^3$ degrees of freedom. 

Hence for $\ell \gg \ell_c$ the minimal surface is determined by the AdS geometry of the infrared CFT and we expect an area law.  In order to get a feeling of the transition from the volume law for $\ell \leq \ell_c$ to the area law for
$\ell \gg \ell_c$, we can continue the analysis with the restricted hypersurface, the capped cylinder, but now with
expression (\ref{cilent}) appropriately matched to an AdS$_7$-like space.  To perform this matching, we consider the
entropy contribution of a `cap' of boundary volume $|A|$ at height $u_H = T_H$ in the AdS space and demand that this
equals (\ref{lstcap}) at $z_* = z_H$. The corresponding entropy associated to a surface capped at $u=u_*$ and extending
up to the matching point $u=u_H =T_H$ is
\begin{equation}
S_{\rm AdS} (u_*) = 2C N^3 \,|A|\,u_*^5 + 2C N^3 \,|\pt A| \,\int_{u_*}^{u_H} {du \over u} \,u^4 \;.
\end{equation}
and the total entropy results from adding (\ref{cyll}) to this expression, evaluated at $z_* = z_H$.  A local minimum
occurs at $u_* = 2/5\ell$, provided $u_* \leq T_H$. In other words, the minimal hypersurface selects a standard local
UV/IR correspondence for $\ell \geq \ell_H = 4\pi R /5$. In the remaining interval $\ell_c < \ell < \ell_H$ the minimum
surface sits at the entrance of the tube, with $u_* = T_H$ and satisfying area law.\footnote{The UV-finite contribution in this case satisfies a volume law.  It is the short-distance contribution with explicit cutoff dependence that follows an area law. In keeping with our emphasis on the UV behavior in this paper, we shall determine the area/volume scaling only in terms of the leading UV contribution to the entanglement entropy. } 

\begin{figure}
  \begin{center}
    \epsfig{file=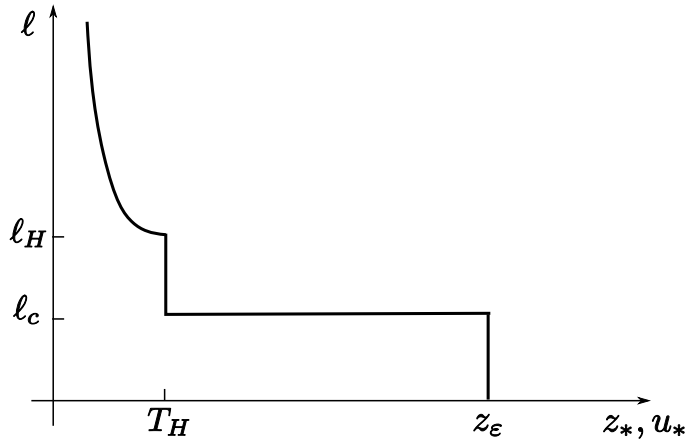, width= 9cm}
    \caption{Schematic plot of the UV/IR relation, as determined by the toy minimal hypersurfaces of capped cylinders
      for a given value of $\ell$, versus the location of the turning point in the bulk. For $\ell > \ell_H$ we have the
      standard `Heisenberg-type' relation $\ell(u_*) \sim 1/u_*$, characteristic of local theories. In the interval
      $\ell_c < \ell < \ell_H$ the minimal surface is stuck at $u_* = u_H = T_H$ (the IR end of the tube). At
      $\ell=\ell_c$ there is a degenerate set of minimal surfaces with turning points anywhere in the tube, and finally
      for $\ell < \ell_c$ the only minimal surface is the one set at the cutoff scale $z=z_\vep$. }  \label{fig:uvirlstl}
  \end{center}
\end{figure}

\begin{figure}
  \begin{center}
    \epsfig{file=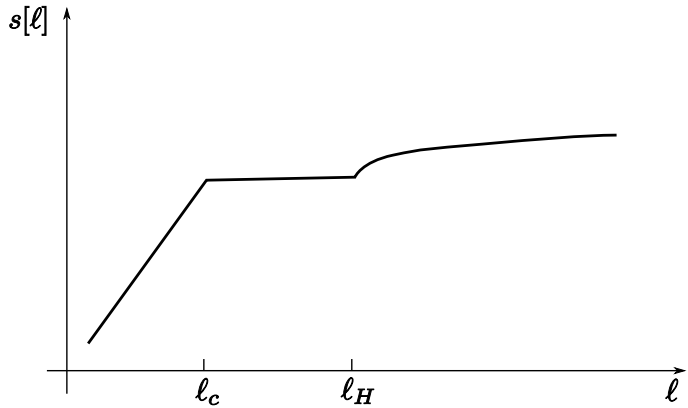, width= 9cm}
    \caption{Schematic plot of the entropy density $s[\ell] = S[A] / |\pt A|$ for the toy minimal hypersurfaces of
      capped cylinders showing the local regime at $\ell \geq
      \ell_H$, the nonlocal volume law at $\ell < \ell_H$ and the intermediate transient.}
  \label{fig:entropyal}
\end{center}
\end{figure}

We summarize the results of this section in Figs.~\ref{fig:uvirlstl} and \ref{fig:entropyal}. The strip entropy density
$s[\ell] = S[A] / |\pt A|$ scales linearly with $\ell$ up to the critical length scale $\ell_c$ according to the volume
law (\ref{volaf}) \beq\label{denv} s[\ell] \sim {N_{\rm eff} (E_\vep) \over \ell_c^{\,5}} \,\ell \;, \qquad \ell <
\ell_c\;.  \eeq 
A short area-law plateau follows
 \beq\label{plat} s[\ell] \sim {N_{\rm eff} (E_\vep)
  \over \ell_c^{\,4}}  \;, \qquad {\rm for}\;\;\ell_c < \ell < \ell_H\;, \eeq and finally we get a very slow growth at
large $\ell$, corresponding to the infrared CFT: \beq\label{infdd} s[\ell] \sim {N_{\rm eff}
  (E_\vep) \over \ell_c^4} \left(1+b \left(1-\ell_H^4 / \ell^4 \right)\right)\;,\qquad {\rm for} \;\;\ell >\ell_H\;,
\eeq where $b$ is a very small constant of $O(E_H / E_\vep)$. Notice that there is no regime in which the
cutoff-dependent terms adopt a field theoretical form. Instead, we find that $\ell_c$ takes the role of effective UV
cutoff in the theory. However, the local regime, with a Heisenberg dispersion $u_* \sim 1/\ell$, is associated with an
{\it area law}, while the nonlocal region is associated to a {\it volume law}.  The sharp transition shown in Fig. \ref{fig:uvirlstl} is expected to be an artifact of our usage of non-smooth hypersurfaces, and should be replaced by
a rapid crossover in the exact treatment. 

\subsubsection{Deconstructed LST}

\noindent

We have seen that the entanglement entropy for strips with small widths ($\ell \lesssim R$) is associated with surfaces lying at
the UV cutoff of LST's dual geometry. We can ask what would happen if the LST model is given a more standard UV completion. For example, we may embed the LST theory into some UV fixed point admitting an AdS description in the gravity regime. In such models, the LST behavior is reduced to some transient in the energy variable or, in the geometric language, to some intermediate `tube-like' geometry interpolating between and infrared (IR) AdS and some UV AdS corresponding to the asymptotic CFT at high energies.  Embeddings of this type can be found in the literature, using ideas of `deconstruction'  \cite{deconst, motl, dorey}. 

\begin{figure}[t]
  \begin{center}
    \epsfig{file=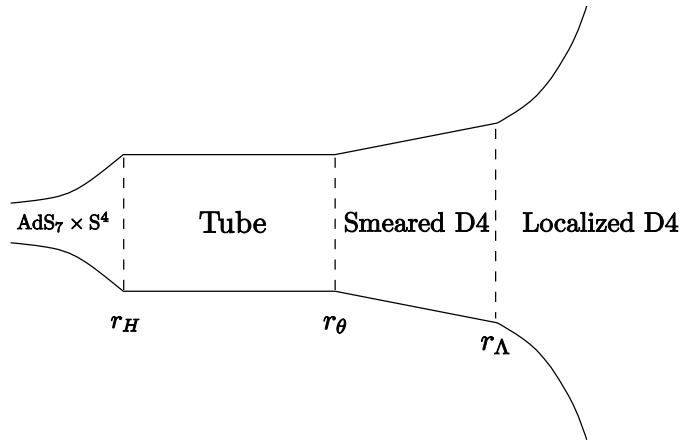, width=3.5truein}
    \parbox{14truecm}{\caption{ Schematic picture of the background profile in IIA
        deconstruction, showing the different regions of interest in the vicinity of the
        LST regime.}
    \label{fig:profiledecons}}
  \end{center}
\end{figure}

One particularly simple model that admits an explicit bulk geometrical description was introduced in \cite{dorey} and recently discussed at length in \cite{littlehag} (see this reference for more details). In this set up the UV fixed point is given by a $(2,0)$ CFT in six dimensions compactified on a circle. The merger with an intermediate LST-like background (\ref{cilingro}) is achieved via two intermediate transients described in Fig. \ref{fig:profiledecons}. 

To be more precise, the  $\mathds{R}^{5+1}$ world-volume of the NS5-brane is compactifed down to $\mathds{R}^{4+1} \times {\rm S}^1$ on a circle of length $L$, with a differential warping between the $\mathds{R}^{4+1}$ and ${\rm S}^1$ factors in such a way that the metric is asymptotic to  that of ${\hat N}$ D4-branes {\it smeared} over the circle of length $L$, where
${\hat N} \sim N^{3/2} L /g_s R$. The associated near-horizon metric
\begin{equation}\label{smeardf}
    ds^2 \approx  \frac{r}{R} \left(-dt^2 + dx^2_4 \right) + \frac{R}{r} \left( dw^2 + dr^2 + r^2 \,d\Omega_3^2  \right) \;, \qquad e^\phi \approx  g_s  
\end{equation}
matches the tube (\ref{cilingro}) at $r\sim r_\theta = R $. The $w$ coordinate parametrizes the circle of
size $L$.
 At even larger radii, of order $r\sim r_\Lambda = L$,  the smeared D4-branes are revealed as an infrared approximation to the metric of ${\hat N}$ localized D4-branes, a system studied in the previous section of this paper.  To achieve the matching one proceeds as in the example around Eq. (\ref{defhache}), defining now $\rho^2 = w^2 + r^2$ as the appropriate radial variable for the localized D4-branes throat. 
 
 Finally, the  D4-branes develop strong coupling and match by an $11^{th}$ dimensional uplift to an ${\rm AdS}_7 \times {\rm S}^4$ background similar to the one appearing in the IR, but associated to a CFT with ${\hat N}^3 $ degrees of freedom in the UV.

\begin{figure}
  \begin{center}
    \epsfig{file=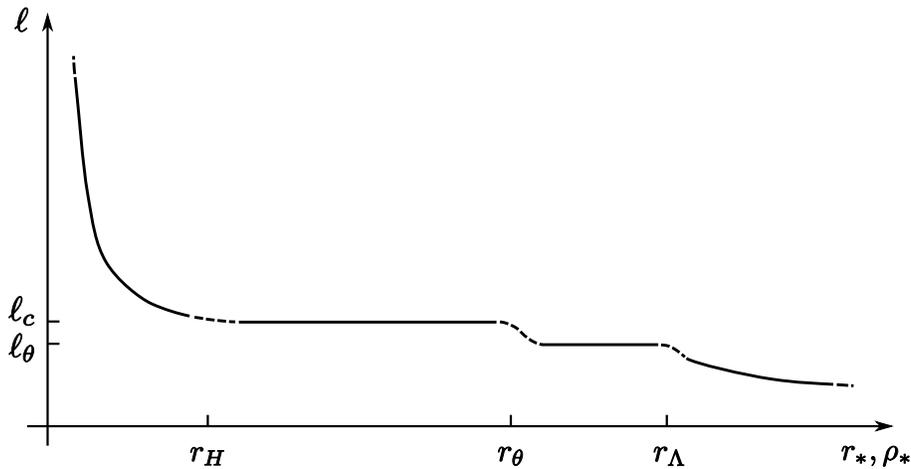, width= 12cm}
    \caption{Schematic plot of $\ell$ as a function of $r_*$, $\rho_*$ for the deconstructed LST background. In the tube
      region $r_H < r < r_\theta$ one has a constant behavior of $\ell$ as well as in $r_\theta < r < r_\Lambda$, whereas
      for the corresponding regions of $\ell > \ell_c$ and $\ell < \ell_\theta$ one has $\ell \sim
      \rho_*^{-\frac{1}{2}}$.}
    \label{fig:deconstructedLST}
  \end{center}
\end{figure}

Let us consider a strip of the form $[-\ell/2 , \ell/2] \times {\rm S}^1_L \times \mathds{R}^3$ and define the entanglement entropy density $s[\ell]$ by factoring out the volume of the $\mathds{R}^3$ factor.  For turning points in the regime described by (\ref{smeardf}), corresponding to $r_\theta \ll r_* \ll r_\Lambda$,  we have an entropy functional 
\begin{equation}\label{ents}
  s[\ell] = \frac{ L \Omega_3}{16 \pi^6 \alpha'^4 g_s^2}  \int_{-\frac{\ell}{2}}^{\frac{\ell}{2}} dx\, r^3 \sqrt{1  +
    \frac{R^2}{r^2} \left(\frac{dr}{dx}\right)^2}
\;.\end{equation}
The smooth extremizing hypersurface then {\it fixes} the strip length to a constant value $\ell(r_*) = \ell_\theta$, independent of $r_*$,
\begin{equation}
  \ell(r_*) =  2 r_*^3 R \int_{r_*}^{\infty} \frac{dr}{r\sqrt{r^{6} - r^{6}_*}} = \frac{ \pi }{ 3 } R\;, 
\end{equation}
just as in the case of the LST tube. The critical length $\ell_\theta$ is somewhat smaller than $\ell_c = \pi R/2$, but with the same order of magnitude.  In this situation, the volume of the bulk hypersurfaces at $\ell = \ell_\theta$ will be
approximately minimized by the one with the largest possible value of $r_*$, i.e.~$r_* \sim r_\Lambda$, the point where the metric is matched to that of localized D4-branes. For $\ell < \ell_\theta$ the turning point will occur inside the standard D4-brane metric, yielding standard
Heisenberg dispersion $\ell \sim 1/u_*$,  for an appropriate energy variable in the D4-brane throat. The resulting
entropy will  show the scaling (\ref{dpeen}) with the replacements $p\rightarrow 4$,  $N \rightarrow {\hat N}$ and $\lambda \rightarrow g_s {\hat N} \sqrt{\alpha'}$.  At even lower values of $\ell$ we enter the six-dimensional CFT scaling.
The qualitative behavior of the dispersion relation is shown in Fig. \ref{fig:deconstructedLST}. 

At any rate, if the ultraviolet cutoff is taken all the way to the region dominated by the UV fixed point, the leading
short-distance behavior of the entropy is guaranteed to be given by the six-dimensional area law
\beq\label{sixa}
s[\ell] \sim {\hat N}^3 {L \over \vep^4}\;,
\eeq
with finite-$\ell$ corrections that will be sensitive to the different thresholds visible in the UV/IR relation.  
The previous volume law is shifted to a volume law of just the UV-finite part of the entanglement entropy.

\subsection{Noncommutative Yang--Mills}
\label{sec:noncommutative}
\noindent

Compared to the example of LST, noncommutative theories epitomize a milder notion of non-locality. Consider a maximally
supersymmetric $SU(N)$ super Yang--Mills theory quantized on a spacetime $\mathds{R}^2_\theta \times \mathds{R}^{1+1}$,
where $\mathds{R}^2_\theta$ is the noncommutative plane defined by a Moyal algebra $[x,y] = i\theta$. 
 Perturbative excitations behave as gluons with a rigid
transversal length $L(p) = \theta p_\theta$, where $p_\theta$ is the projection of the momentum onto the noncommutative
plane. The presence of these `rigid rod' degrees of freedom introduces a basic nonlocality in the theory by the
corresponding violation of Lorentz invariance, but it does not strictly affect the number of local degrees of freedom.

While propagation of such extended gluons is not affected by the noncommutative deformation, nontrivial $\theta$
dependence only arises in the interacting theory at the level of nonplanar corrections in the $1/N$ expansion. In
particular, the density of states at large $N$ is not sensitive to the noncommutative deformation.

The dual holographic description of these theories was introduced in Refs. \cite{noncads}, using the basic scaling of
\cite{adscft} in the string theory set up of Ref. \cite{seibergwitten}.  The metric is \beq\label{ncads} ds^2/R^2 = u^2
\left(-dt^2 + dz^2 + f(u) (dx^2 + dy^2) \right) + {du^2 \over u^2} + d\Omega_5^2 \;, \eeq with a dilaton and
Neveu--Schwarz B-field: \beq\label{ncdil} e^{2\phi} = g_s^2 f(u)\;,\qquad B_{xy} = {1\over \theta} (1-f(u))\;, \eeq
where $g_s$ is the asymptotic string coupling in the infrared region $u\rightarrow 0$, related to the Yang--Mills
coupling constant by $g_{\rm YM}^2 = 2\pi g_s$. The curvature of the AdS region is controlled by the usual expression
$R^4 = 4\pi g_s N \alpha'^2$, and the profile function \beq\label{prof} f(u) = {1\over 1+ (a_\theta u)^4}\;, \qquad
a_\theta = \sqrt{\theta} \,(4\pi g_s N)^{1/4} =(2\lambda)^{1/4} \,\sqrt{\theta} \;, \eeq determines the
$\theta$-dependence through the effective length scale $a_\theta \propto \sqrt{\theta}$, renormalized by a fractional
power of the 't Hooft coupling, a common occurrence in AdS holographic duals. In this form, the model is clearly
asymptotic to the standard AdS$_5 \times {\rm S}^5$ background at small values of $u$, which gives the energy coordinate
of the infrared fixed point.

There is a further subtlety regarding the proper interpretation of this model which is of some relevance for our
discussion below. The induced metric on the boundary, obtained as usual removing the $u^2 R^2$ factor at fixed $u$, has
in this case an anisotropy caused by the presence of the $f(u)$ factor in the noncommutative plane coordinated by
$(x,y)$. It is important however to realize that the physically relevant metric to which the energy-momentum tensor of
the noncommutative theory couples is the so-called `open-string metric', defined in Ref.~\cite{seibergwitten} as
\beq\label{opens} G_{ij} = g_{ij} - ( \alpha'_{\rm eff})^2 \left( B \;{1\over g}\; B \right)_{ij} \;, \eeq where
$\alpha'_{\rm eff}$ is the effective string slope parameter and $g_{ij}$ is the metric entering the string sigma-model.
In the case of the metric induced at fixed $u$ by (\ref{ncads}) we have (restricting to the noncommutative plane)
$g_{ij} = f \delta_{ij}$, $B_{ij} = \theta^{-1} (1-f) \delta_{ij}$ and the effective string tension can be obtained by
dropping a fundamental string at fixed $u$. Its mass per unit length is
$$
{1\over 2\pi\alpha'_{\rm eff}} = {R^2 u^2 \over 2\pi \alpha'} \;,
$$
which determines $\alpha'_{\rm eff}$. Using $a_\theta^8 = \theta^2 R^4 /\alpha'^4$ from their definitions, we finally
obtain $G_{ij} = \delta_{ij}$, i.e.~the physical metric of the noncommutative theory is the standard Euclidean metric,
despite the deformation induced by the holographic background \cite{liwu} (for a recent example where this subtlety
makes all the difference, see \cite{karl}).  This means that, when considering the areas and volumes of a prescribed
region, we will define $|A|$ and $|\pt A|$ as coordinate areas and volumes, using the standard Euclidean metric on
$\mathds{R} \times \mathds{R}^2_\theta$, rather than the induced metric as it comes from (\ref{ncads}). Conversely, the
bulk volume that enters the holographic ansatz of the entanglement entropy will be computed in the bulk metric.

\subsubsection{The computation}

\noindent

Let us consider the strip of {\it coordinate} width $\ell$ as entanglement region, and define $s[\ell]$ as the entropy
density resulting from factorizing out the longitudinal volume of $\mathds{R}^2$. The behavior of the entanglement
entropy is very sensitive to the orientation of this $\mathds{R}^2$ plane of the strip, since the system has lost
Lorentz invariance by the $\theta$ deformation. It is easy to see that the entropy functional is $\theta$-independent
when the strip plane is parallel to the noncommutative plane $\mathds{R}^2_\theta$.  Hence, the results coincide with
those of the standard CFT in that case.  In all other possible orientations, one finds a nontrivial result. We shall
consider as representative the orthogonal orientation, in which the strip plane is orthogonal to
$\mathds{R}^2_\theta$. Without loss of generality we can align the strip along the $y$ direction, so that $\ell$ is the
coordinate extent of the strip in the $x$ direction. Then, the entropy functional takes the form \beq\label{entrn}
s[\ell] = {|{\rm S}^5| R^8 \over 32\pi^6 \alpha'^4 g_s^2} \int_{-\ell/2}^{\ell/2} dx \,u^3 \sqrt{1+ {(du/dx)^2 \over u^4
    f(u)}}\;, \eeq for a straight belt defined by a function $u(x)$ with turning point at $u_* = u(0)$, determined by
the equation \beq\label{turning} \ell(u_*) = 2 u_*^3 \int_{u_*}^\infty {du \over u^2 \sqrt{f(u) (u^{6} - u_*^{6})} } =
{2\over u_*} \int_1^\infty {ds \sqrt{1+ (a_\theta u_*)^4 s^4} \over s^2 \sqrt{s^6 -1}}\;.  \eeq

\begin{figure}
  \begin{center}
    \epsfig{file=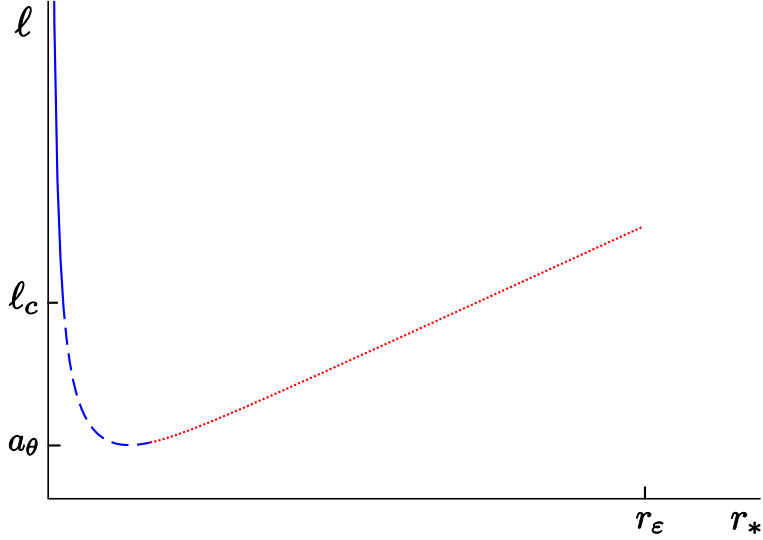, width= 10cm}
    \caption{Numerical plot of the UV/IR relation in the noncommutative theory. Stable hypersurfaces (blue continuous line) disperse as $\ell
      \sim 1/u_*$ and unstable ones (red dotted line) disperse as $\ell \sim u_*$. The hypersurfaces with $\ell <
      \ell_c$ in the blue dashed line are metastable. Notice that there are no extremal smooth
      3-surfaces with $\ell < \ell_{\rm min} \sim a_\theta $.} 
    \label{fig:noncommutativelu}
  \end{center}
\end{figure}

This function is shown in Fig.~\ref{fig:noncommutativelu}. It has a minimum at $\ell =\ell_{\rm min} \sim a_\theta$ and
implies that there are no extremal, smooth hypersurfaces for $\ell < \ell_{\rm min}$. Conversely, for $\ell> \ell_{\rm
  min}$ there are two extremal hypersurfaces of which only the one with lower value of $u_*$ is a local minimum of the
entropy functional.  In the deep infrared $u_* a_\theta \ll 1$ we can approximate (\ref{turning}) by the usual local
UV/IR relation, \beq\label{stanr} \ell(u_*) \approx {c_0 \over u_*}\;, \qquad c_0 = 2\int_1^\infty {ds \over s^2
  \sqrt{s^6 -1}} = 2 \sqrt{\pi} \frac{\Gamma\left( \frac{2}{3} \right)}{\Gamma\left( \frac{1}{6} \right)} \;.  \eeq On
the other hand, in the deep noncommutative regime $u_* a_\theta \gg 1$ we have an exotic dispersion relation for the unstable
extremal surfaces.  \beq\label{nonst} \ell(u_*) \approx c_\infty \,a_\theta^2 \,u_* \;, \qquad c_\infty = 2\int_1^\infty
{ds \over \sqrt{s^6 -1}} = \frac{\sqrt{\pi}}{3}\frac{\Gamma\left( \frac{1}{3} \right)}{\Gamma\left( \frac{5}{6} \right)}\;.  \eeq

\begin{figure}
  \begin{center}
    \epsfig{file=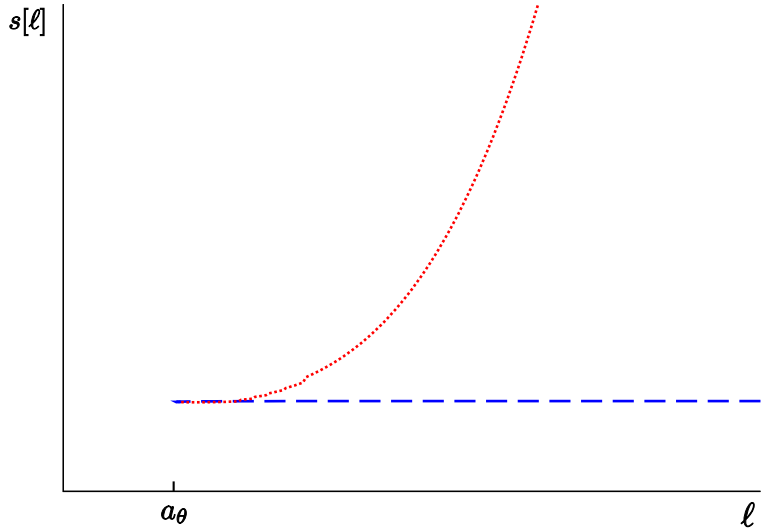, width= 10cm}
    \caption{Numerical plot of the entanglement entropy for the smooth extremal hypersurfaces.  The blue dashed line
      represents the metastable solutions whereas the red dashed line gives the entropy of the unstable solutions.}
    \label{fig:noncommutativeslsmooth}
  \end{center}
\end{figure}

The entropy functional   evaluated at the stable solution 
\beq\label{lead}
s[\ell] = N^2 \,{\Omega_5 \over \pi^4} \int_{u_*}^{u_\vep} {du \,u^4 \over \sqrt{f(u)(u^6 - u_*^6)}} 
\eeq
has a leading short-distance behavior
\beq\label{shortd}
s[\ell] \sim N^2 \,a_\theta^2 \, u_\vep^4 \;,
\eeq
with the $\ell$-dependent contribution being of order $-N^2 /\ell^2$ and thus small in the limit of very large $u_\vep$. We can compare this to the entropy of the degenerate surface sitting  at the cutoff scale, $u=u_\vep$,  which scales extensively and is independent of $f(u)$, 
\beq\label{uvs}
s[\ell]_{\rm UV} \sim N^2 \ell \, u_\vep^3\;.
\eeq
We see that (\ref{uvs}) is smaller than (\ref{shortd}) provided $\ell < \ell_c$ with
\beq\label{critcl}
\ell_c \sim {u_\vep a_\theta^2} \sim  {a_\theta^2 \over \vep} \sim {\theta \over \vep} \sqrt{\lambda}\;,
\eeq
where $\lambda =g_{\rm YM}^2 N$ is the 't Hooft coupling of the IR fixed point and we have defined an effective cutoff length $\vep \sim u_\vep$ (see Fig. 4). That $u_\vep$ is the standard energy coordinate even for 
$u_\vep a_\theta \gg 1$ is guaranteed by the known fact that the Hawking temperature of a black hole in the noncommutative bulk geometry (\ref{ncads}) is independent of $\theta$, as well as the Bekenstein--Hawking entropy. Hence, the $u$-coordinate of the horizon measures the temperature of the plasma phase of the NCYM theory, at least in the planar approximation.

\begin{figure}
  \begin{center}
    \epsfig{file=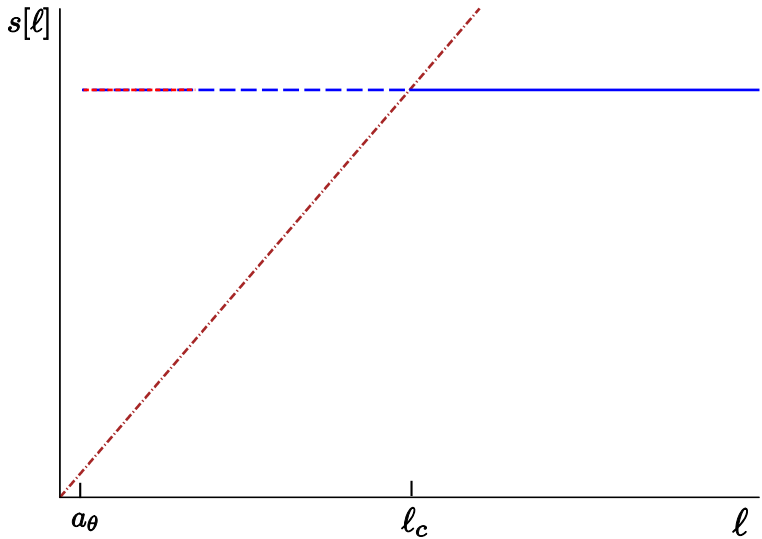, width= 10cm}
    \caption{Numerical plot showing how the smooth stable 3-surface ceases to be the absolute minimum of induced volume
      at $\ell =\ell_c$. It is replaced by the extensive one at the cutoff scale for $\ell < \ell_c$ (brown
      dashed-dotted line).}
    \label{fig:noncommutativeslall}
  \end{center}
\end{figure}

Therefore, we can summarize the situation as follows. For large values of  the strip's width, $\ell \gg \ell_c$, the short-distance contribution to the entanglement entropy shows an area law of the form 
\beq\label{shof}
S[A] \propto N_{\rm eff} {|\pt A | \over \vep^2} \;, \qquad N_{\rm eff} = N^2 \left({\ell_c \over a_\theta}\right)^2\;,
\eeq
in particular, we see a renormalized value of the effective number of degrees of freedom. In this respect, the noncommutative theory differs markedly from its commutative counterpart. It should be stressed that only the cutoff-dependent contribution sees a deformed number of degrees of freedom as in (\ref{shof}), since the finite $\ell$-dependent part has the standard scaling 
$$
C[\ell] \equiv  \ell {d\over d\ell}\,s[\ell] \sim {N^2 \over \ell^2} \;, \qquad {\rm for}\;\;\;\ell \gg \ell_c\;.
$$
 This behavior changes abruptly at $\ell \sim \ell_c$, and we switch to  a {\it volume law}, characteristic of the nonlocal regime
\beq\label{nonex}
S[A] \propto N^2  {|A| \over \vep^3} \;,
\eeq
where this time $N_{\rm eff}$ is given by the standard IR value, $N^2$.

Finally, let us briefly point out that theories with noncommutative time can be formally defined, with $[t,z] = i
\theta_e$. Even though these models are plagued with a variety of consistency problems \cite{nctimebunch}  one can still carry out 
the analysis of the entanglement entropy  at a formal level, along the preceding lines.  The holographic dual  has the same structure as (\ref{ncads}) with the replacement of the `magnetic' $(x,y)$ plane by the `electric' $(t,z)$ plane in the metric, dilaton and $B$-field profiles. The $(x,y)$  warping $f(u)$ is substituted by $(t,z)$ warping  $f_e (u)$, obtained by the replacement  $\theta \rightarrow \theta_e$  and $ a_\theta \rightarrow a_e$.

Repeating the previous analysis for this particular background we find that the local UV/IR relation 
$\ell u_* \sim 1$ holds in order of magnitude for the case that the strip length $\ell$ extends along the $x$ or $y$ directions, even for $\ell a_e \ll 1$. Accordingly, the area law holds for any $\ell > \vep$. The short-distance scaling of the entanglement entropy still reveals an exotic number of degrees of freedom $N_{\rm eff} \sim N^2 (a_e /\vep)^2$, just as in the case of `magnetic' noncommutativity.  
For the case that the strip length $\ell$ extends along the $z$ axis, one finds a critical length $\ell_c \sim a_e^2 /\vep$ for the transition to a volume law, just as the magnetic case. Now however the effective number of degrees of freedom jumps from $N_{\rm eff} \sim N^2 (\ell_c / a_e)^4 $ at large $\ell$ to
$N_{\rm eff} \sim N^2 (\ell_c / a_e)^2$ at  $\ell < \ell_c$.   

\subsubsection{Interpretation of UV/IR mixing}

\noindent

A key consequence of our analysis is the occurrence of very strong UV/IR mixing effects in the noncommutative
theory. The naive scale of nonlocality is $\sqrt{\theta}$, or rather its strong-coupling version $a_\theta$. However, we
see that the effects on the entanglement entropy, i.e.~the onset of the volume law, take place at a length scale
$\ell_c \sim a_\theta^2 /\vep$, larger than the naive one by a factor of $a_\theta /\vep$.

Continuing this result to the weakly coupled theory, it would mean that the effective scale of nonlocality is $\theta
/\vep$ rather than $\sqrt{\theta}$.  In fact, this turns out to have a rather natural explanation. As reviewed in the
introduction to this subsection, the elementary excitations of NCYM in perturbation theory are a sort of extended gluons,
which behave as rigid rods of transverse size $L_{\rm eff} (p) \sim \theta p_\theta$, with $p_\theta$ the projection of
the momentum onto the noncommutative plane. Since the maximum momentum is $1/\vep$, we find that the maximum size of a
gluon is a rod of length $\theta /\vep$ in the $\mathds{R}^2_\theta$ plane. This reproduces the expected effective scale
of nonlocality, and also explains why a strip which is cut parallel to the noncommutative plane is insensitive to this
`growth' of the gluons.

\section{Epilogue: Lorentz symmetry, entanglement entropy and the density of states}
\label{sec:epilogue}
\noindent

The two examples studied in this paper might induce in the reader the impression that models with
UV volume law are not that difficult to construct. In fact, we would like to argue that these two models, LST and NCYM, are quite special. We have already emphasized that the noncommutative model owes  its volume law to the occurrence of rigid extended objects, particularly breaking Lorentz symmetry. In this section we point out that keeping Lorentz symmetry in the boundary theory severely restricts the  possibilities. 

More specifically, we will consider bulk systems with Einstein-frame metric of the form
\beq\label{efgen}
ds^2/R^2 = \lambda(u)^2 (-dt^2 + dx_d^2 \,) + {du^2 \over \mu(u)^2}\;,
\eeq
where the warp factors $\lambda(u), \mu(u)$ give the most general metric compatible with Lorentz symmetry on the $\mathds{R}^{d+1}$ boundary theory. We can also assume that  $\lambda(u) > 0, \mu(u) >0$ and that $\lambda(u), \mu(u) \rightarrow u$ as $u\rightarrow 0$, i.e.~we have an IR fixed point with
$N_{\rm eff} \sim R^d /G_{d+2}$ effective degrees of freedom.\footnote{In fact, we can relax this condition and keep some thresholds at low $u$ related to nontrivial IR phenomena, such as mass gaps and confinement. Since we are emphasizing here the UV behavior, those details will not affect our analysis.}   This family of metrics includes LST, all near-horizon brane metrics and flat space as particular cases, but excludes noncommutative models with explicit violation of Lorentz symmetry. 

We are interested in the behavior of minimal hypersurfaces at very large $u$. Using again the simple {\it ansatz} of a capped cylinder of base $\pt A$ reaching down to $u=u_m$, we have
\beq\label{cila}
S(u_m) \sim N_{\rm eff} |A|\,\lambda(u_m)^{d} + N_{\rm eff} |\pt A|\, \int_{u_m}^{u_\vep} {du\over \mu(u)} \lambda(u)^{d-1}\;,
\eeq
where the first term arises from the cap of geometry $|A|$ located at $u=u_m$ and the second term is the volume of the cylinder reaching out from $u_m$ up to the cutoff $u_\vep$. The turning point $u_*$ is obtained by extremizing this expression with respect to $u_m$, leading to
\beq\label{exg}
\ell(u_*) \sim {1\over \mu(u_*) \lambda'(u_*)}\;,
\eeq
as the modified UV/IR relation, where $\lambda'(u)$ is the derivative of the warp factor with respect to $u$. Thus we recover the standard Heisenberg-like relation for the conformal case $\lambda(u) \sim \mu(u) \sim u$. 

We can define a theory with {\it volume law} in the UV by requiring that the expression (\ref{cila}) is minimal at the UV cutoff, i.e.~one does not decrease the total volume by lowering the position of the cap in the bulk spacetime. This condition is
\beq\label{volcon}
\mu(u_\vep) \lambda' (u_\vep) < {c\over \ell}\;,
\eeq
where $c$ is a constant of $O(1)$ and we evaluate the profile factors at the UV cutoff $u_\vep$ to indicate that we are interested in the deep UV behavior of the metric. 

Now we can relate this behavior to the density of states of the theory, as defined by the Bekenstein--Hawking entropy of black holes with planar horizon. The corresponding black metrics take the form
\beq\label{blacefgen}
ds^2/R^2 = \lambda(u)^2 (-h(u) dt^2 + dx_d^2 \,) + {du^2 \over \mu(u)^2 h(u)}\;,
\eeq
with  $h(u)$ a Schwarzschild-like factor with a first-order zero at the location of the horizon, $h(u_0) =0$,
with $h'(u_0) \sim 1/u_0$.  Using standard methods we get for the Hawking temperature and Bekenstein--Hawking entropy density over $\mathds{R}^d$:
\beq\label{termo}
T(u_0) = {\lambda(u_0) \mu(u_0) \over b \,u_0} \;, \qquad s(u_0)_{\rm bh} \sim N_{\rm eff} \,\lambda (u_0)^d\;,
\eeq
where $b$ is a positive constant of $O(1)$. With the standard definition of the running effective number of degrees of freedom (species degeneracy)  we have
\beq\label{runefff}
N_{\rm eff} (u_0) \equiv {s(u_0) \over T^d} \propto N_{\rm eff} \left({u_0 \over \mu(u_0)}\right)^d\;.
\eeq
We  will say that a model is `well behaved' when the running species degeneracy {\it does not decrease} as we access higher energies, i.e.~$dN_{\rm eff} /du_0 \geq 0$. A further condition satisfied by a `decent'
holographic dual is that the specific heat should be positive, i.e.~$dT/du_0  =T' (u_0) \geq 0$.
  Taking now the derivative of the temperature function we derive the expression
$$
\mu(u_0) \lambda' (u_0) = u_0 \,b\, T' (u_0) + b\,T(u_0) -  \lambda(u_0) \mu' (u_0)\;,
$$
and the last two terms can be related to the derivative of the running effective number of degrees of freedom, so that we can finally write
\beq\label{monot}
\mu(u_\vep) \lambda' (u_\vep) = b \,u_\vep\,T' (u_\vep) + b\,d\,T(u_\vep)\,u_\vep\,{d\log N_{\rm eff} \over du} {\Big |}_{u=u_\vep}\;.
\eeq
  Hence, we see that a positive specific heat $T'(u_\vep) >0$ in the UV, combined with a non-decreasing species degeneracy  essentially guarantees that the inequality (\ref{volcon}) will be violated and 
 the entanglement entropy
 {\it will not} satisfy a volume law in the UV. In other words, we will see an area law, because the UV asymptotics will be dominated by the cylinder rather than the cap.  
 
 Field-theoretical densities of states have a powerlike growth of $T(u_0)$,
 which is enough to ensure area law, even for an asymptotically constant $N_{\rm eff} (u_0)$.  The effect of $N_{\rm eff}$ in the argument is much milder, since any powerlike growth of $N_{\rm eff} (u_0)$ only yields a constant
 $d\log N_{\rm eff} / d\log u$ and a corresponding constant term on the right hand side of (\ref{monot}). 
 
 Conversely, models with Lorentz invariance on the boundary and volume law must have a `pathological' density of states,
 either because the specific heat is negative, or because the species degeneracy decreases at high energies.  For
 example, we may consider the case of flat space, with $\lambda(u) = \mu(u) = 1/R$, whose holographic dual, if formally
 defined, is expected to be a nonlocal theory \cite{marolf}. Conforming to these expectations, when one calculates the
 entanglement entropy one finds it satisfies the volume law. And indeed, the density of states of black holes has an
 effective temperature $T(u_0) \sim (R^2 u_0)^{-1}$ with negative specific heat. The formal dimensional reduction of a
 higher-dimensional flat space behaves in a similar fashion, as well as the D$p$-brane metrics with $p \ge 5$: they all
 present a volume law for the entanglement entropy and again both have negative specific heat and shrinking number of
 species (in verifying these examples, it is important to notice that (\ref{efgen}) is written in Einstein-frame
 conventions after dimensional reduction to $d+2$ dimensions).

 It is interesting to notice that the LST model is precisely a marginal case from the point of view of this analysis,
 since  the effective temperature is constant $T=T_H$ in the `LST
 plateau'.  On the other hand, the NCYM model evades the discussion in this section, due to the violation of Lorentz
 symmetry, since the directional distortion of the bulk metric cancels out when computing both the Hawking temperature
 and the Bekenstein--Hawking entropy of black holes.

\section{Conclusions}
\label{sec:conclusions}
\noindent

In this paper we have strengthened the basic intuition that a certain degree of nonlocality tends to introduce a volume
law in the scaling of the entanglement entropy, as opposed to the more standard area law, characteristic of local
QFT. We have done this at very strong coupling, using the holographic definition of entanglement entropy, and testing
these ideas in the case of two models with an available geometrical description, namely Little String Theory and
noncommutative super Yang--Mills theory. 

Our results are also interesting probes into the peculiar workings of holography in these nonlocal theories. Both models
have standard IR fixed points with an AdS/CFT description and an intrinsic length of nonlocality. We find in both cases
that the volume-law entanglement entropy measures the effective number of degrees of freedom at high energies, weighed
by the same number of degrees of freedom that get exposed by highly excited thermal states.

Both models pose interesting challenges beyond the leading classical approximation in the bulk description. In the case
of LST, it has been emphasized recently that string loop corrections tend to destabilize the Hagedorn density of states,
unless maximal energy cutoff is in place \cite{littlehag}. In the case of NCYM it is well known that non-planar
corrections bring on the UV/IR effects into full strength \cite{minwalla}. Since one of the most important open problems
in the holographic theory of entanglement entropy is the generalization beyond the classical approximation, these models
will represent very stringent checks on any proposal in this direction.

Finally, we have seen that the two models studied in this paper have a rather peculiar status. One can argue that the
combination of Lorentz symmetry plus a more or less standard density of states at high energy is sufficient to guarantee
an area law in the UV contribution to the entanglement entropy. The LST model arises as a marginal, exceptional case in
this analysis, whereas the nocommutative model evades the argument by the violation of Lorentz symmetry. Not
surprisingly, extending this treatment to the case of the holographic dual of strings in flat space, suspected to
be a highly non-local theory, one finds a volume law of the entanglement entropy, thus endorsing the interpretation of
volume-law scaling as a criterion of non-locality.

\noindent

\subsubsection*{Acknowledgements}

\noindent

C.~A.~F.~wishes to thank J.~I.~Cirac for pointing out the spin-chain example quoted in the introduction and G.~Sierra
for useful discussions.  This work was partially supported by MEC and FEDER under grant FPA2006-05485, the Spanish
Consolider-Ingenio 2010 Programme CPAN (CSD2007-00042), Comunidad Aut\'onoma de Madrid under grant HEPHACOS P-ESP-00346
and the European Union Marie Curie RTN network under contract MRTN-CT-2004-005104. C.~A.~F.~enjoys a FPU fellowship from
MEC under grant AP2005-0134.


\end{document}